%% file: main.tex
\documentclass[reprint,
amsmath,amssymb,
aps,
floatfix,
superscriptaddress,
]{revtex4-2}

\usepackage[export]{adjustbox}[2011/08/13]
\usepackage{graphicx} 
\usepackage{dcolumn} 
\usepackage{colordvi}
\usepackage{color}
\usepackage{pstricks}
\usepackage{epstopdf}
\usepackage{amssymb}
\usepackage{url}
\graphicspath{{ps}}
\usepackage{hyperref}
\usepackage{tabularx}
\usepackage{multirow}
\usepackage{siunitx}
\usepackage{hyphenat}
\usepackage{subfigure}
\usepackage{graphicx,float}
\usepackage[italic]{hepnames}
\usepackage{color, colortbl}
\usepackage{mathtools}
\usepackage{orcidlink}

\newcommand{\Bpipi}{\ensuremath{\overline{B}{}^0 \to \pi^0 \pi^0}\xspace}

\input ./belle2sym.tex

\begin{document}

\def\belletwo {\it {Belle II}}

\vspace*{-3\baselineskip}
\resizebox{!}{3cm}{\includegraphics{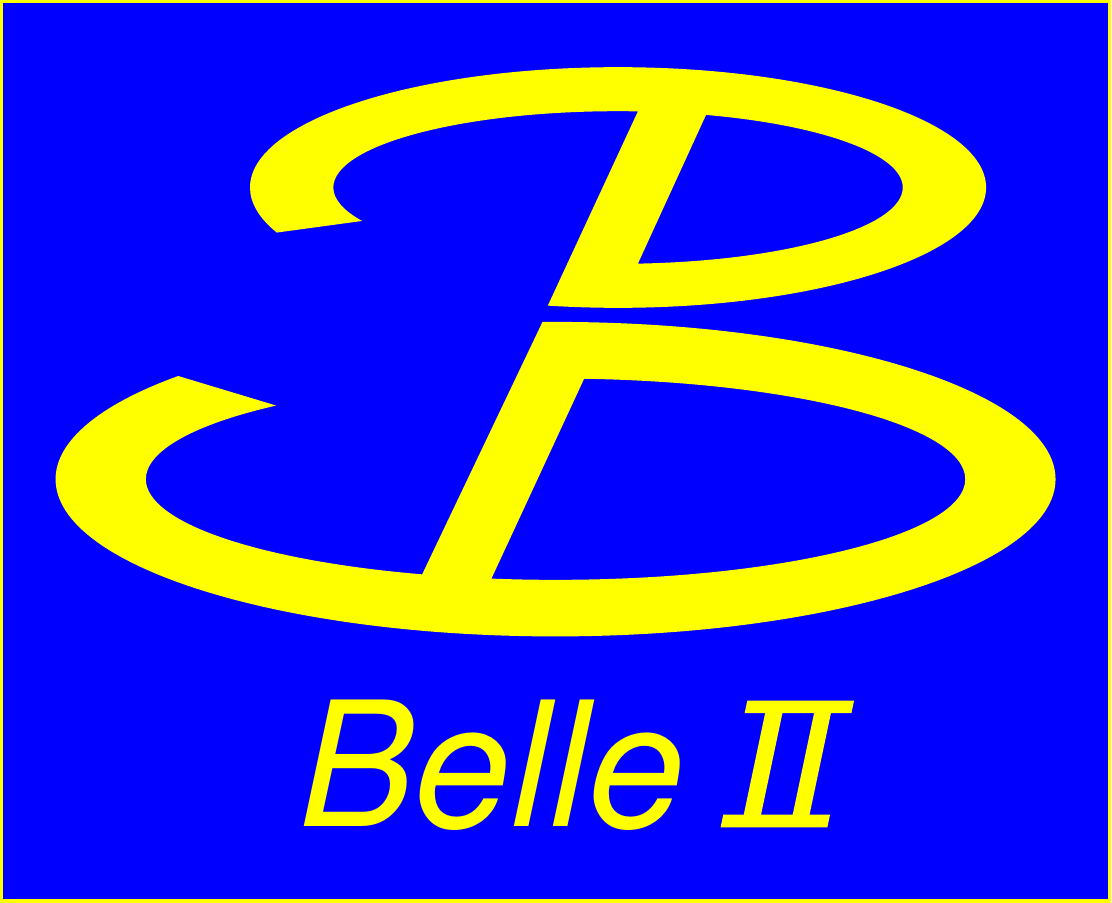}}

\vspace*{\baselineskip}
\begin{flushright}

\end{flushright}

\vspace{1.5cm}
\title {Measurement of the branching fraction and $\it CP$ asymmetry of $B^{0} \rightarrow \pi^{0} \pi^{0}$ decays using $198 \times 10^6$ $B\overline{B}$ pairs in Belle II data}

\input{authors-orcid}

\begin{abstract}

We report measurements of the branching fraction and $\it CP$ asymmetry in $B^{0} \to \pi^{0} \pi^{0}$ decays reconstructed at Belle II in an electron-positron collision sample containing $198 \times 10^{6}$ $B\overline{B}$ pairs. We measure a branching fraction $\mathcal{B}(\Bpipi) = (1.38 \pm 0.27 \pm 0.22) \times 10^{-6}$ and a $\it CP$ asymmetry $\Acp(\Bpipi) = 0.14 \pm 0.46 \pm 0.07$, where the first uncertainty is statistical and the second is systematic. 

\end{abstract}


\maketitle

{\renewcommand{\thefootnote}{\fnsymbol{footnote}}}
\setcounter{footnote}{0}

\newpage

The study of decay-time-dependent $\it CP$ asymmetries in decays dominated by the $\mbox{\ensuremath{b \to u}}$ transition, specifically hadronic decays of bottom mesons into charmless two-body final states, is currently the most precise way to measure the least known angle of the unitarity triangle, $\phi_2$ (or $\alpha$) $\equiv \textrm{arg}\left( -V_{td} V^*_{tb} / V_{ud} V^*_{ub} \right)$. Here, $V_{ij}$ are elements of the Cabibbo-Kobayashi-Maskawa (CKM) quark-mixing matrix~\cite{Kobayashi:1973fv}. Improved measurements of $\phi_2$ will test the unitarity of the CKM matrix and constrain possible flavor-structure extensions of the standard model (SM). One approach is to measure the time-dependent decay-rate asymmetry between $\overline{B}{}^0$ and $B^0$ mesons that decay to $\pi^+\pi^-$ final states. This asymmetry would be proportional to sin$(2\phi_2)$ if the decay involved only tree-level $b \rightarrow u$ processes. However, the asymmetry is affected by an unknown and difficult-to-predict shift with respect to the desired $\phi_2$ angle due to the presence of $b \rightarrow d$ loop (`penguin') contributions. The tree-level and penguin amplitudes have similar magnitudes, so the shift is sizable and complicates the determination of $\phi_2$. The penguin and tree contributions can be disentangled using the $B \to \pi \pi$ isospin relations~\cite{Gronau1990, Charles:2017evz}
\begin{equation}
     A^{+0} = \frac{1}{\sqrt{2}}A^{+-} + A^{00} \quad \textrm{and} \quad \bar{A}^{-0} = \frac{1}{\sqrt{2}}\bar{A}^{+-} + \bar{A}^{00},
\end{equation}
where $A^{ij}$ and $\bar{A}^{ij}$ are amplitudes for the decays $B \to \pi^i \pi^j$ and $\overline{B} \to \pi^i \pi^j$, respectively. Here, $B$ and $\pi$ indicate charged or neutral bottom-mesons and pions, respectively, while $i$ and $j$ refer to electric charge. Taking advantage of these relations requires precise measurements of the branching fraction $\mathcal{B}$ and ${\it CP}$ asymmetries of each $B \rightarrow \pi \pi$ decay mode. The greatest limitation to exploiting the isospin relations lies in the uncertainty of the $B^0 \to \pi^0 \pi^0$ inputs, $\mathcal{B}$ and the time-integrated $\it CP$ asymmetry,
\begin{equation}
    \Acp = \frac{\Gamma(\overline{B}{}^0 \to \piz \piz)-\Gamma(B^0 \to \piz \piz)}{\Gamma(\overline{B}{}^0 \to \piz \piz)+\Gamma(B^0 \to \piz \piz)},
    \label{eq:acp}
\end{equation}
where $\Gamma$ is the decay width. The world-average values $\mathcal{B}(B^0 \to \pi^0 \pi^0) = (1.59 \pm 0.26) \times 10^{-6}$ and $\mathcal{A}_{\it CP}(B^0 \to \pi^0 \pi^0) = 0.33 \pm 0.22$~\cite{ParticleDataGroup:2022pth} combine measurements reported by the BaBar~\cite{PhysRevD.87.052009} and Belle~\cite{Belle:2017lyb} collaborations. Additional measurements would improve our knowledge of $\phi_2$.

Theoretical predictions for $\mbox{\ensuremath{\mathcal{B}(B^0 \rightarrow \pi^0 \pi^0)}}$ based on QCD factorization~\cite{Beneke:1999br,Beneke:2003zv,Beneke:2005vv,Pilipp:2007mg} and perturbative QCD~\cite{Lu:2000em, Zhang:2014bsa} are approximately five times smaller than the world average value. Furthermore, the ratio of color-suppressed to color-allowed tree amplitudes, as inferred from other charmless two-body decay modes, does not agree well with expectations~\cite{Charng:2004ed}. This might indicate large electroweak-penguin contributions, which are difficult to explain in the SM~\cite{Mishima:2004um, Charng:2003iy}. Various approaches, which predict a wide range of values for $\mathcal{B}$ and $\Acp$, have been proposed as possible solutions to this disagreement~\cite{Qiao2015, Li2011, Li2017, Cheng2015}. More precise measurements of these quantities would help in discriminating among the various solutions proposed to address this discrepancy. In addition, a better understanding of the color-suppressed tree amplitude could help resolve the so-called $\B \to K\pi$ puzzle~\cite{Li:2005kt,Baek:2009pa,Barger:2004hn}

In this paper, we present a measurement of $\mathcal{B}$ and \Acp for the $\Bpipi$ decay using a data sample consisting of $(198.0 \pm 3.0) \times 10^{6}$ $B\overline{B}$ pairs~\footnote{Throughout this paper, charge-conjugate modes are implicitly included unless noted otherwise.} collected from 2019 through 2021~\cite{Belle-II:2019usr}. The sample is collected with the Belle II detector, located at the SuperKEKB asymmetric-energy $e^+e^-$ collider~\cite{Akai:2018mbz}. A full description of the Belle II detector is given in Ref.~\cite{Belle-II:2010dht}. The detector consists of several subdetectors arranged in a cylindrical structure around the beam pipe. The $z$ axis of the lab frame is defined as the symmetry axis of a superconducting solenoid, which generates a 1.5~T uniform field along the beam direction. The positive direction is given by the electron-beam direction, and the polar angle, $\theta$, is defined with respect to the +$z$ axis. The detector is divided into three regions, and in increasing order of $\theta$, they are referred to as the forward endcap, barrel, and backward endcap. The inner subdetectors are a silicon pixel detector surrounded by a four-layer double-sided silicon strip detector and a central drift chamber (CDC). These subdetectors are used to reconstruct charged particles and measure their momentum. A time-of-propagation counter~\cite{Wang:2017ajq} and an aerogel ring-imaging Cherenkov detector cover the barrel and forward endcap regions, respectively, and are used for charged particle identification. The electromagnetic calorimeter (ECL) is a segmented array of 8736 thallium-doped cesium iodide [CsI(Tl)] crystals arranged in a projective geometry toward the interaction point and covering about 90\% of the solid angle in the center-of-mass (c.m.) frame. The ECL identifies electrons and photons in an energy range of 20 MeV to 4 GeV and occupies the remaining volume inside the superconducting solenoid. Resistive plate chambers and scintillating fibers to identify muons and $K^0_L$ mesons are installed in the flux return of the magnet. 

We use GEANT4-based~\cite{GEANT4:2002zbu} simulated samples to optimize event selection criteria, compare distributions observed in data with expectations, determine fit models, calculate signal efficiencies, and study sources of background. To study the signal, we use $10^{7}$ $\Y4S \to \B^0 \overline{B}{}^0$ decays generated with EvtGen~\cite{Lange:2001uf}, where one $B$ meson decays as $B^0 \to \piz \piz$. To study backgrounds, we use a simulated sample approximately five times larger than the data sample. This sample consists of $\epem \to \Y4S \to B\overline{B}$ processes and continuum $\epem \to \qqbar$ background, generated with EvtGen and PYTHIA~\cite{Sjostrand:2014zea} where $q$ denotes a $u$, $d$, $s$, or $c$ quark. To account for a large observed $\tau^+ \tau^-$ background, we use a sample of $\epem \to \tau^+\tau^-$ events generated with KKMC~\cite{Jadach:1999vf} and TAUOLA~\cite{Jadach:1993hs} that is the same size as the continuum sample. To validate our analysis, we use the $B^{0} \to \overline{D}{}^0(\to K^+ \pi^- \pi^0) \pi^0$ decay as a control mode, as it contains two \piz particles in the final state and has an order of magnitude more yield. We use a simulated sample of $5 \times 10^{6}$ control-mode events generated with EvtGen. To calibrate and validate our photon reconstruction, we use the $D^{*+} \rightarrow D^0(\rightarrow K^0_S (\to \pi^+ \pi^-) \pi^0) \pi^+$ mode. The data are processed with the Belle II analysis software framework~\cite{Kuhr:2018lps, basf2-zenodo}. This is the first measurement of this channel at Belle II.

Measuring \Bpipi decay properties is challenging, as the decay is both CKM-suppressed and color-suppressed. As the final state consists of photons with no tracks, it is difficult to reconstruct. In addition, the large number of neutral pions produced in $e^+e^- \to q\bar{q}$ continuum events can be combined to mimic the $\Bpipi$ signal. Finally, the reconstruction is susceptible to extraneous photons arising from beam interactions with the beam pipe and residual gas; this is referred to as beam-induced background. Hence, conventional and machine-learning based approaches, validated on data, are employed to achieve optimized selections. The signal yield and $\Acp$ are determined by performing a maximum likelihood fit to the data. 

The online-event selection requires all events to pass criteria based on total energy and neutral-particle multiplicity. In the offline analysis we identify photon candidates by requiring that the number of crystals in an ECL energy deposition (cluster), which can be fractional as a result of energy splitting with nearby clusters, be greater than 1.5. The cluster timing is required to be within 200 ns of the offline estimated event time. We require that the cluster energy exceed 20.0~MeV in the barrel region of the ECL ($32.2^{\circ} < \theta < 128.7^{\circ}$), and 22.5~MeV in the forward endcap ($12.4^{\circ} < \theta < 31.4^{\circ}$) and backward endcap ($130.7^{\circ} < \theta < 155.1^{\circ}$) regions. Since scintillation light from CsI(Tl) crystals has a relatively long
decay time, high-energy events from Bhabha processes ($e^+e^- \to e^+e^-$) can overlap with subsequent hadronic event signals. A random photon from the hadronic event can be combined with the residual energy (misreconstructed photon) in the CsI(Tl) crystals to form a \piz candidate. To suppress non-signal photons, and to account for the angular dependence of ECL-related variables, we employ a boosted decision tree (BDT)~\cite{Keck:2017gsv}, separately for each of the three polar-angle regions of the ECL. The BDT is trained on a simulated sample of \Bpipi events that includes the effect of beam-induced background and uses ten input variables: the photon energy and transverse momentum, the energy recorded in the crystal having the highest signal, the distance between the ECL cluster and the nearest charged particle hitting the ECL, four variables that depend on how energy is distributed between the clusters, and two variables that depend on the fraction of cluster energy detected in the central crystal. We refer to this classifier as the ``Photon-BDT". We impose a requirement on the Photon-BDT output that maximizes a figure-of-merit $\it{S}/\sqrt{\it{S+B}}$, where $\it{S}$ and $\it{B}$ are the expected number of signal $B^0$ and background events, respectively. In simulated samples, this selection removes 68\% of misreconstructed photons and retains 96\% of genuine photons. Studies of the $D^{*+} \rightarrow D^0(\rightarrow K^0_S (\to \pi^+ \pi^-) \pi^0) \pi^+$ decay are used to validate the Photon-BDT classifier's performance. The signal efficiency and Photon-BDT output distribution are consistent between simulations and data. 

Selected photons are paired to form \piz candidates. We require that the \piz momentum in the lab frame be greater than 1.5 \gevc, and that the angle between the momenta of final-state photons in the lab frame be less than 0.4 radians. These requirements suppress the combinatorial background from low-energy photons. The cosine of the helicity angle, defined as the angle between the higher energy $\gamma$ direction in the $\piz$ rest-frame and the $\piz$ direction in the lab frame, is required to be less than 0.99 to reject misreconstructed $\piz$'s, which tend to peak very close to one. The diphoton mass is required to be between 0.115 $\gevcc$ and 0.150 $\gevcc$, which corresponds to a range of approximately $+2.0 \sigma$ and $-2.5 \sigma$ about the known \piz mass. The mass requirement is asymmetric as the reconstructed \piz mass has a slight negative skew due to energy leakage from the ECL calorimeter. We improve the momentum resolution of the $\pi^0$ candidates by performing a kinematic fit that constrains their mass to the known value~\cite{ParticleDataGroup:2022pth}. Signal $B^0$ candidates are reconstructed by combining two $\piz$ candidates. To select signal $B^0$ candidates, two kinematic variables are defined,
\begin{equation}
    \mbc = \sqrt{E_{\rm beam}^2 - |\vec{p}_{B}|^2} \quad \text{and} \quad \dele = E_{B} -  E_{\rm beam},
\end{equation}
where $E_{\rm beam}$ is the beam energy and $(E_{B},\vec{p}_{B})$ is the reconstructed four-momentum of the $B$ candidate. All quantities are calculated in the c.m.\ frame of the \Y4S resonance. The $\mbc$ and $\dele$ distributions of signal decays peak at the $B$ mass and zero, respectively. Candidate $B$ mesons are required to have $5.26 < \mbc < 5.29~\gevcc$ and $-0.3 < \dele < 0.2~\gev$. The $\dele$ requirement is not centered around zero because of energy leakage from the ECL cluster. 

In data, photon-energy corrections are applied to correct for ECL miscalibration. Studies of the $\mbox{\ensuremath{D^{*+} \rightarrow D^0(\to K^0_S (\pi^+ \pi^-) \pi^0) \pi^+}}$ control mode are used to validate the corrections. The \piz momentum is predicted using the momenta of the charged pions and energy-momentum conservation. In simulation, this predicted momentum is typically closer to the true momentum than the measured momentum as the momentum of charged pions is measured more precisely in the CDC than photon energies are measured in the ECL. When the corrections are applied, the data-simulation difference between the predicted and measured \piz momentum decreases, and the difference between the $\dele$ peak position in data and simulation is approximately 1~MeV. 

The sample includes a large continuum background. To reduce this background, we use topological variables that take advantage of the jet-like nature of \qqbar events and the spherical distribution of $B\overline{B}$ events. We train a BDT classifier to analyze 28 variables comprising modified Fox-Wolfram moments~\footnote{The Fox-Wolfram moments were introduced in \href{https://doi.org/10.1103/PhysRevLett.41.1581}{G. C. Fox and S. Wolfram, Phys. Rev. Lett. 41, 1581 (1978)}. The modified moments used in this paper are described in \href{https://doi.org/10.1103/PhysRevLett.87.101801}{K. Abe et al. (Belle Collaboration) Phys. Rev. Lett. 87, 101801 (2001) and \href{https://doi.org/10.1016/S0370-2693(01)00626-8}{K. Abe et al. (Belle Collaboration), Phys. Lett. B 511, 151 (2001)}.}}, sphericity-related quantities~\cite{PhysRevD.1.1416}, thrust-related quantities~\cite{Farhi:1977sg}, and sets of concentric cones with various opening angles centered around the thrust axis. Variables that show correlations with $\dele$ and $\mbc$ greater than 5\% are excluded. We train the continuum classifier to identify statistically significant signal and background features using simulated signal samples and sideband data; the latter consists of events that satisfy all selection criteria but are in a signal-depleted region $5.22 < \mbc < 5.27 \gevcc$ and $0.1 < \dele < 0.5 \gev$. We use these simulated samples to determine a minimum threshold $C_{\min}$ of the continuum classifier output $C$ that minimizes the expected statistical uncertainty of the \Acp measurement. This selection rejects 93\% of the background while retaining 76.5\% of the signal. The continuum classifier output is transformed into a Gaussian-like shape according to $T_{c} = \text{log}[(C - C_{\min})/({C_{\max} - C})]$, where $C_{\max}$ is the maximum value of the continuum classifier output. Candidate $B$ mesons are required to have $|\Tc| < 3$. The \Tc distributions of signal candidates and continuum are expected to peak at one and zero, respectively. 

After suppression of continuum background, 1.3\% of events have more than one $B^0$ candidate. For such events, the average multiplicity is 2.03 candidates per event. We choose the candidate with the minimum sum of the absolute deviations of the reconstructed $\piz$ masses from the known value~\cite{ParticleDataGroup:2022pth}. This requirement is 56\% efficient in selecting the correct $B^0$ candidate. Following all selections, 35.5\% of signal events remain, of which 99.0\% are correctly reconstructed. The high fraction of correctly reconstructed events is due to the low percentage of $B\bar{B}$ events in which there are three high momentum $\pi^0$ candidates.

The resulting event sample consists of four main components: signal, continuum, background from non-signal $B$ decays ($B\overline{B}$ background), and $\tau^+ \tau^-$ events. From studies of simulated samples, we find that 90\% of $\B\overline{B}$ background is from $B^+B^-$ events in which the $B^+$ meson decays into a $\rho^+ \pi^0$ final state and the charged pion from the subsequent $\rho^+ \to \pi^+ \pi^0$ decay is not reconstructed. The remaining 10\% is dominated by $B^0 \overline{B}{}^0$ events in which the $B^0$ meson decays into a $K^0_S(\to \pi^0 \pi^0) \pi^0$ final state, and one $\piz$ from the $K^0_S$ decay is not reconstructed. The $B\overline{B}$ background peaks at similar values of $\mbc$ and $T_{c}$ as the signal, but its $\dele$ distribution is shifted to negative values due to the energy lost from the signal-decay daughter that is not reconstructed. In addition, 2.9\% of signal candidates arise from $\tau^+ \tau^-$ events; these candidates are treated as part of the continuum background because their respective $\mbc$, $\dele$, and $\Tc$ distributions are nearly identical. 

\begin{figure*}[th!]
    \makebox[0.329\textwidth]{\includegraphics[clip, trim=0.75cm 0cm 0.2cm 0cm, width=0.333\textwidth]{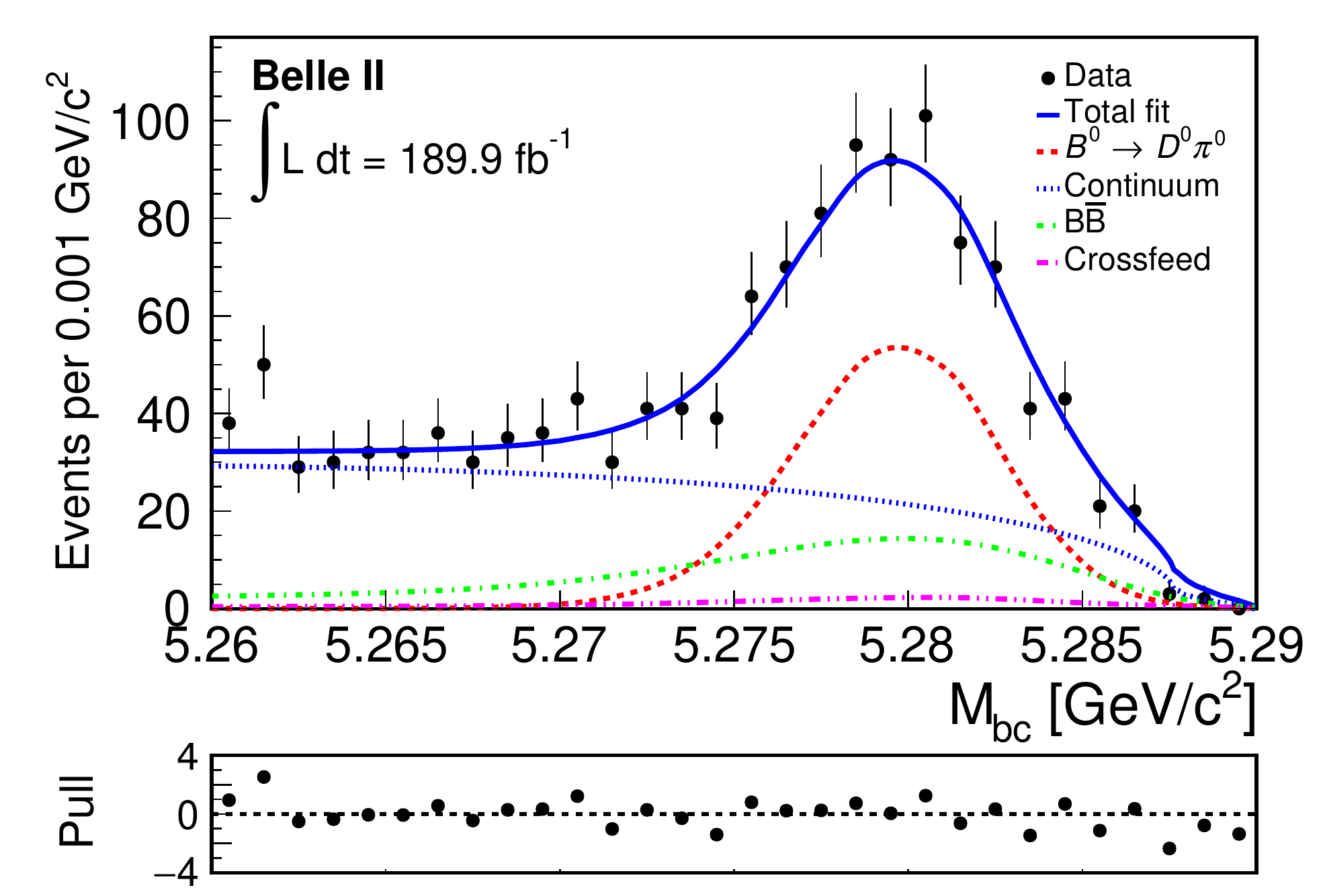}}
    \makebox[0.329\textwidth]{\includegraphics[clip, trim=0.75cm 0cm 0.2cm 0cm,width=0.333\textwidth]{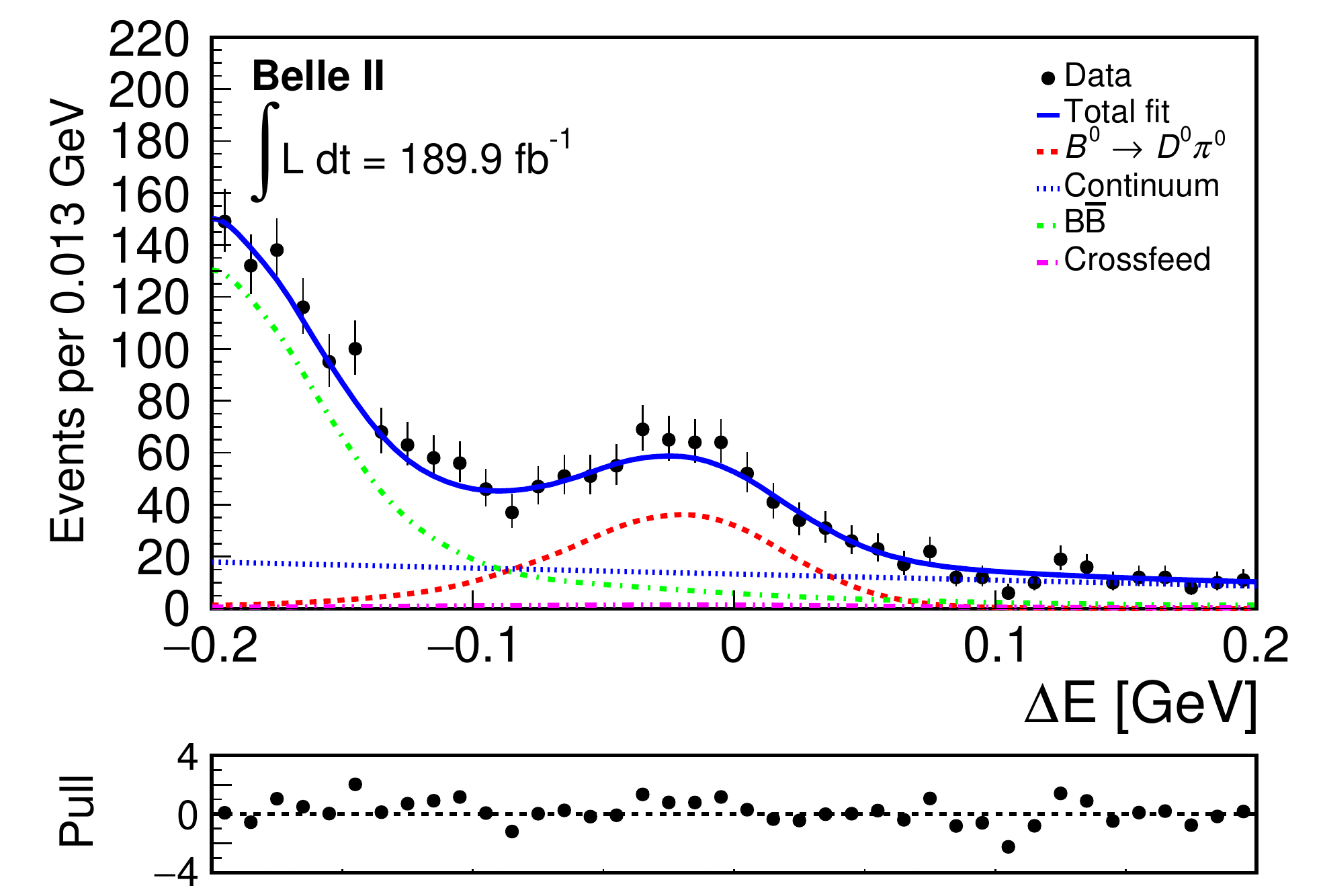}}    
    \makebox[0.329\textwidth]{\includegraphics[clip, trim=0.75cm 0cm 0.2cm 0cm,width=0.333\textwidth]{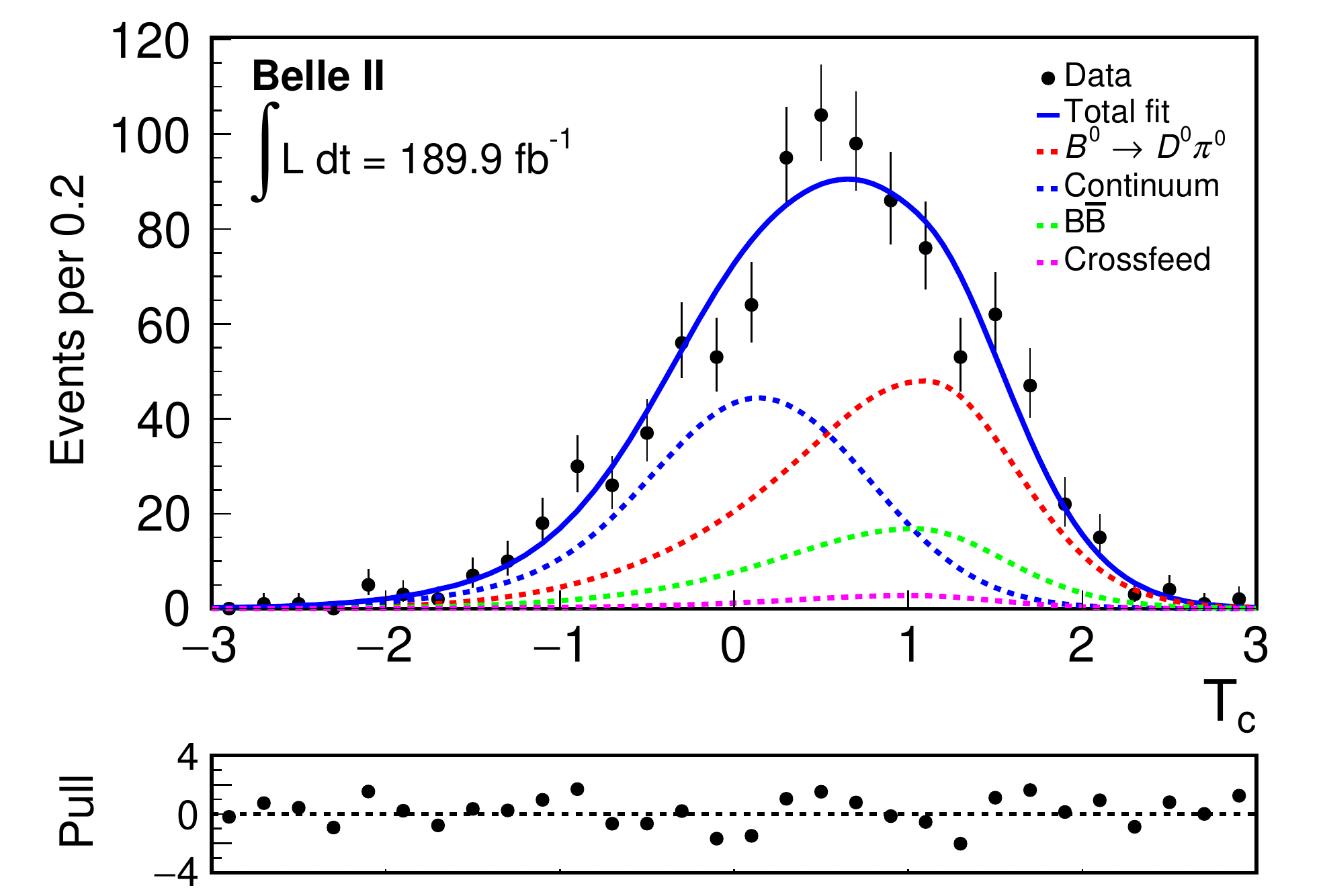}}
    \caption{Distributions of $\mbc$ (left), $\dele$ (middle), and $T_{c}$ (right) for the $B^{0} \rightarrow \overline{D}{}^0(\rightarrow K^+ \pi^- \pi^0) \pi^0$ candidates, for all seven $r$ bins combined. The result of the fit to the data is shown as a solid blue curve. The fit components are shown as a red dashed curve (signal), blue dotted curve (continuum background), green dash-dotted curve ($B\overline{B}$ background), and magenta solid-dotted curve (crossfeed). The plots are signal-enhanced, which correspond to candidates with $5.275 < \mbc <5.285 \gevcc$, $-0.10 < \dele < 0.05 \gev$, and $0 < T_{c} < 3$. When the respective variable is displayed, the selections on that variable are not applied. The difference between observed and fit value divided by the uncertainty from the fit (pulls) are shown below each distribution.}
    \label{fig:control_mode}
\end{figure*}

To measure $\Acp$, the flavor of the signal $B^0$ is determined by reconstructing the accompanying (tag-side) $B$ meson in each event using the category-based algorithm described in Ref.~\cite{Belle-II:2021zvj}. The tagging information is encoded in two parameters: the $b$ or $\bar{b}$ flavor of the tag-side $B$ ($q$) and the purity ($r$). The value $q = +1$ tags a $B^0$, whereas $q = -1$ tags a $\overline{B}{}^0$. The value of $r$ is the algorithm's confidence for an assigned $q$ value. It is defined as $r = 1 -2w$, where $w$ is the fraction of wrongly tagged events and ranges from zero, for no flavour distinction between $B^0$ and $\overline{B}{}^0$, to one for an unambiguous flavor assignment. For example, for a sample of events having $r=0$, there would be an equal number of correctly and incorrectly tagged events; for a sample having $r=1$, there would be no incorrectly tagged events. We divide signal candidates into seven intervals of $r$, with intervals having approximately equal numbers of events. The signal yield and \Acp values are determined by performing a three-dimensional ($\mbc$, $\dele$, $T_{c}$) unbinned extended maximum likelihood fit simultaneously to events in the seven intervals of $r$. The likelihood function is given by
\begin{equation}
\begin{aligned}
& \mathcal{L} = \frac{e^{-\sum_j N^j}}{ \prod_{k} N_k!} \\
    & \times \prod_{k} \left[ \prod_{i=1}^{N_k}\left(\sum_j f^j_k N^j P^j_k(M^i_{\rm bc}, \dele^i, T^i_c, q^i)\right) \right],
\end{aligned}
\end{equation}
where $i$ is the number of candidates, $j$ is the sample component in terms of signal (s), continuum (c), and $B\overline{B}$, and $k$ indicates the $r$ interval. Here, $N^j$ denotes the yield for component $j$, $N_k$ denotes the number of candidates in the $k$th bin, $f^j_k$ is the fraction of candidates in the $k$th bin for the $j$th component, and $P^j_k(M^i_{\rm bc}, \Delta E^i, T^i_c, q^i)$ is the probability density function (PDF) to have the $i$th event of the $j$th component in the $k$th bin. The values of $f^{\rm c}_k$ implicitly include a factor of one-half due to the division of the data into positive and negative $q$ values for each $r$ intervals. Sideband data are used to determine $f^{\rm c}_k$, while $f^{B\overline{B}}_k$ is obtained from large simulated samples.

The PDF for the signal component is
\begin{equation}
\begin{aligned}
P^s_{k}(M_{\rm bc},\Delta E, T_{c},q) = [1-q \Delta w_{k} + q \Delta \epsilon_{k}    (1-2w_{k}) \\
    + [ q (1-2w_{k}) + \Delta \epsilon_{k} (1-q \Delta w_{k})] (1-2\chi_d)\mathcal{A}_{\it CP}] \\
    P^s(M_{\rm bc},\Delta E, T_c),
\end{aligned}
\end{equation}
where $w_{k}$ is the fraction of signal events incorrectly tagged (wrong-tag), $\Delta w_k$ is the difference in the wrong-tag fraction between positive and negative tags, and $\Delta \epsilon_k = \Delta \epsilon_k / 2\epsilon_k$ is the asymmetry of the tagging efficiency. Here, $\epsilon_k$ is the tagging efficiency and $\Delta \epsilon_k$ is the difference in the tagging efficiency between positive and negative tags. The fraction of signal events in each $r$ interval ($f^s_k$), along with $w_k$, $\Delta w_k$, and $\Delta \epsilon_k$, are fixed to values obtained from a fit to $B^0 \to D^{(*)-} h^+$ decays, where $h^+$ stands for a $\pi^+$ or $K^+$, following Ref.~\cite{Belle-II:2021zvj}. The $\it CP$ asymmetry in data is diluted by a factor ($1 - 2w$) due to incorrect tagging, and by a factor of ($1 - 2\chi_{d}$) due to $\BzBzb$ mixing, where $\chi_{d} = 0.1875 \pm 0.0017$ is the time-integrated $\BzBzb$-mixing probability~\cite{ParticleDataGroup:2022pth}. 

The $T_{c}$ PDFs of the signal, continuum, and $B\overline{B}$ components are each modeled using the sum of a Gaussian and a bifurcated Gaussian function with independent mean and width parameters. The \Tc PDFs for the signal and $B\overline{B}$ are modeled independently for each $r$ bin using simulated data; this accounts for an observed dependence of \Tc on $r$. The \Tc PDF for continuum events is the same for all $r$ bins and is taken from the data sideband. 

For the $(\mbc, \dele)$ modeling, a small correlation between $\mbc$ and $\dele$ for the signal is taken into account by using a two-dimensional kernel-density shape. To simplify the $\mbc$ and $\dele$ modeling of the $B^+B^-$ and $\BzBzb$ backgrounds, we assume that they follow the same distributions as the dominant $B^+ \to \rho^+ \pi^0$ and $B^0 \to K^0_S(\to \pi^0 \pi^0) \pi^0$ decays, respectively. The PDFs for the $B\overline{B}$ backgrounds are the sum of two ARGUS functions in $\mbc$ and a kernel-density shape in $\dele$. All signal and $B\overline{B}$ PDF parameters are fixed to those obtained from fits to large samples of simulated events. 

The upper endpoint of the $\mbc$ distribution depends on the beam energy, which varied throughout the course of data taking. To account for this, the continuum is modeled with eight ARGUS functions that have endpoints evenly spaced from 5.287 to 5.290 $\gevcc$. The contribution of each ARGUS function is fixed to the fraction of events reconstructed at each of the corresponding c.m.\ energies. Using eight ARGUS functions models well the variation of the $\mbc$ endpoint and provides a good fit to the data. The $\dele$ distribution of the continuum is modeled with a straight line. We determine the parameters of the continuum PDF for all $r$ bins by fitting to the data sideband region. A small dependence of \dele on $\qr$ found in simulated samples is neglected, as there are insufficient events in the higher $r$ bins of the data sideband for a reliable fit. The \qr distribution for the continuum events shows an asymmetry that could bias the \Acp results. This asymmetry, defined similarly to Eq.(\ref{eq:acp}), is determined to be $0.033 \pm 0.002$. To account for this, we include a \qr asymmetry term in the continuum PDF that is equal to the \qr asymmetry term extracted from the sideband data. From simulated experiments, small biases of 1\% in the branching fraction and 0.02 in \Acp are found; we treat these biases as systematic uncertainties.

The reconstruction and fitting procedure is further validated using $B^0 \to \overline{D}{}^0(\to K^+ \pi^- \pi^0) \pi^0$ decays. This control sample includes a small crossfeed component in which a particle from the tag side is mistakenly included in the signal reconstruction. All photon and $\pi^0$ selections are the same, with the exception of the 1.5 \gevc threshold on $\pi^0$ momentum, which is removed since the $\pi^0$ from the $D^0$ has significantly lower momentum than the $\pi^0$ from a signal decay. We determine the branching fraction to be $\mathcal{B}(B^0 \to \overline{D}{}^0(\to K^+ \pi^- \pi^0) \pi^0) = (3.66 \pm 0.21) \times 10^{-5}$, and the direct $\it CP$ asymmetry to be $\Acp(B^0 \to \overline{D}{}^0 \pi^0) = 0.01 \pm 0.16$. The uncertainties for the control mode measurements are statistical only. These values agree with previous measurements~\cite{Belle:2021nyg}.  Figure~\ref{fig:control_mode} shows signal-enhanced projections of the fits to data. The signal-enhanced region is defined as $5.275 < \mbc <5.285 \gevcc$, $-0.10 < \dele < 0.05 \gev$, and $0 < T_{c} < 3$; for each plot, the selection on the plotted variable is not applied. On average, these signal-enhanced regions contain 47\% of signal decays but only 11\% of background. The control mode is also used to calibrate the $\dele$ width of the signal mode, which is taken from simulation.

\begin{figure*}[th!]
    \makebox[0.329\textwidth]{\includegraphics[clip, trim=0.75cm 0cm 0.2cm 0cm, width=0.333\textwidth]{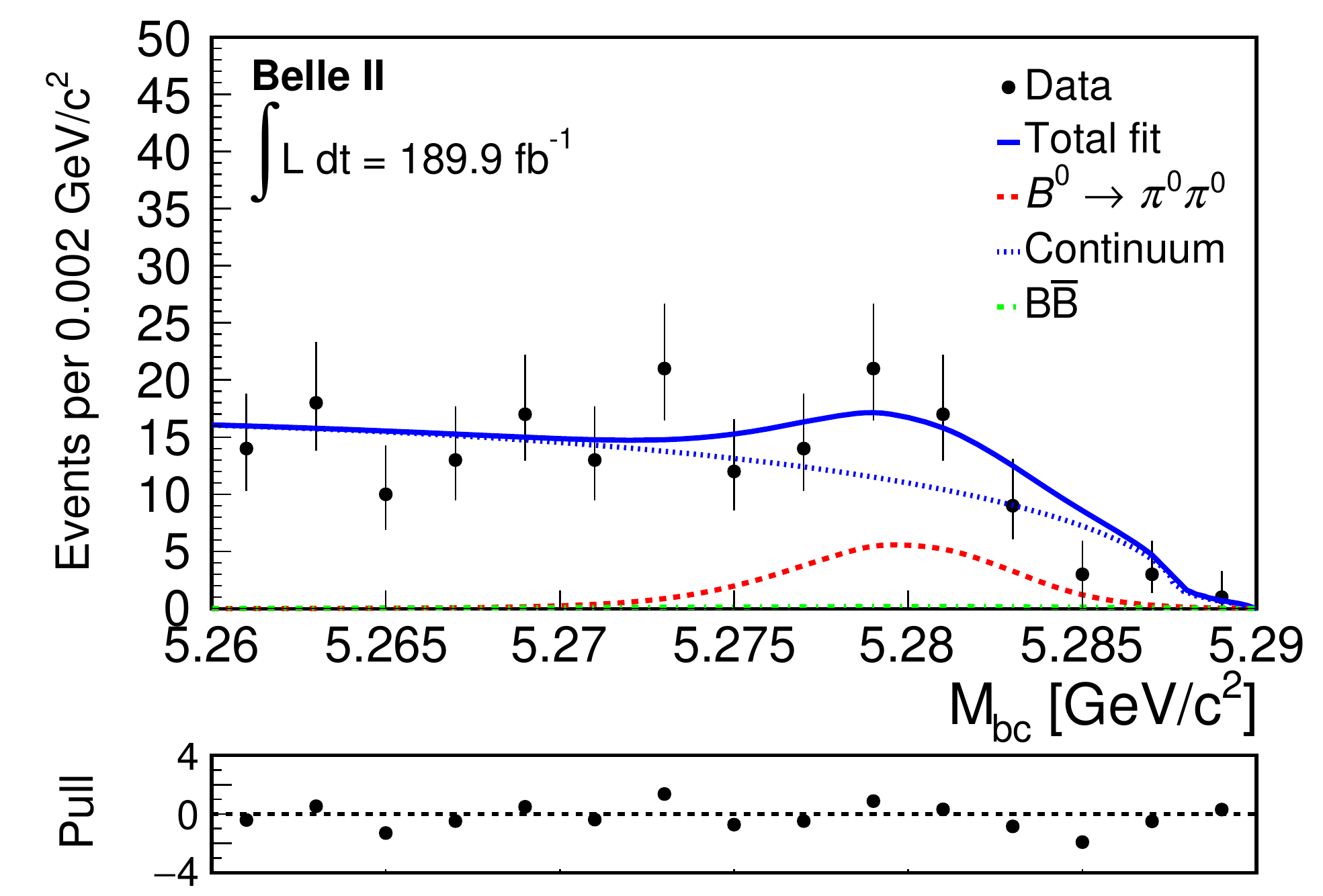}}
    \makebox[0.329\textwidth]{\includegraphics[clip, trim=0.75cm 0cm 0.2cm 0cm,width=0.333\textwidth]{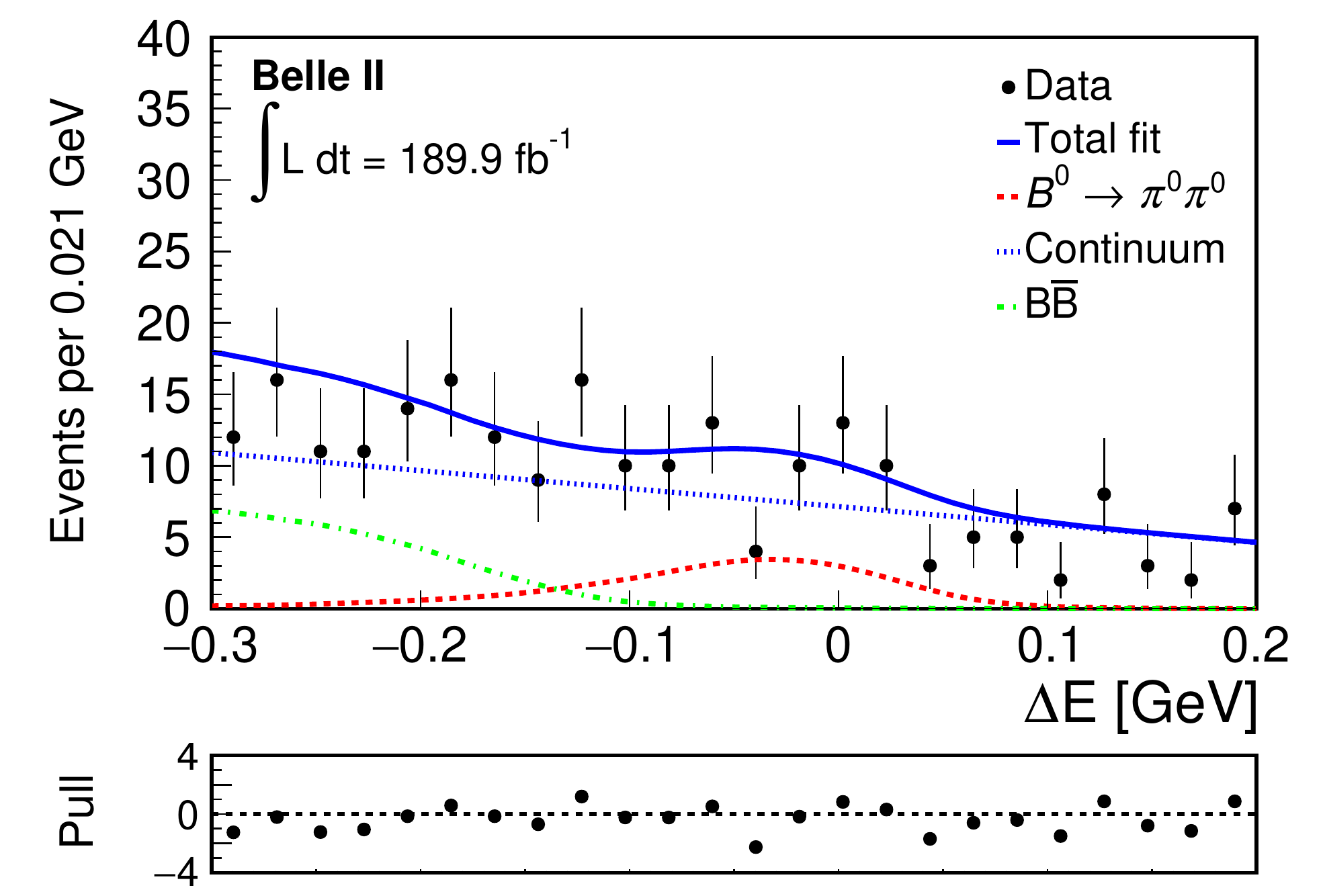}}    
    \makebox[0.329\textwidth]{\includegraphics[clip, trim=0.75cm 0cm 0.2cm 0cm,width=0.333\textwidth]{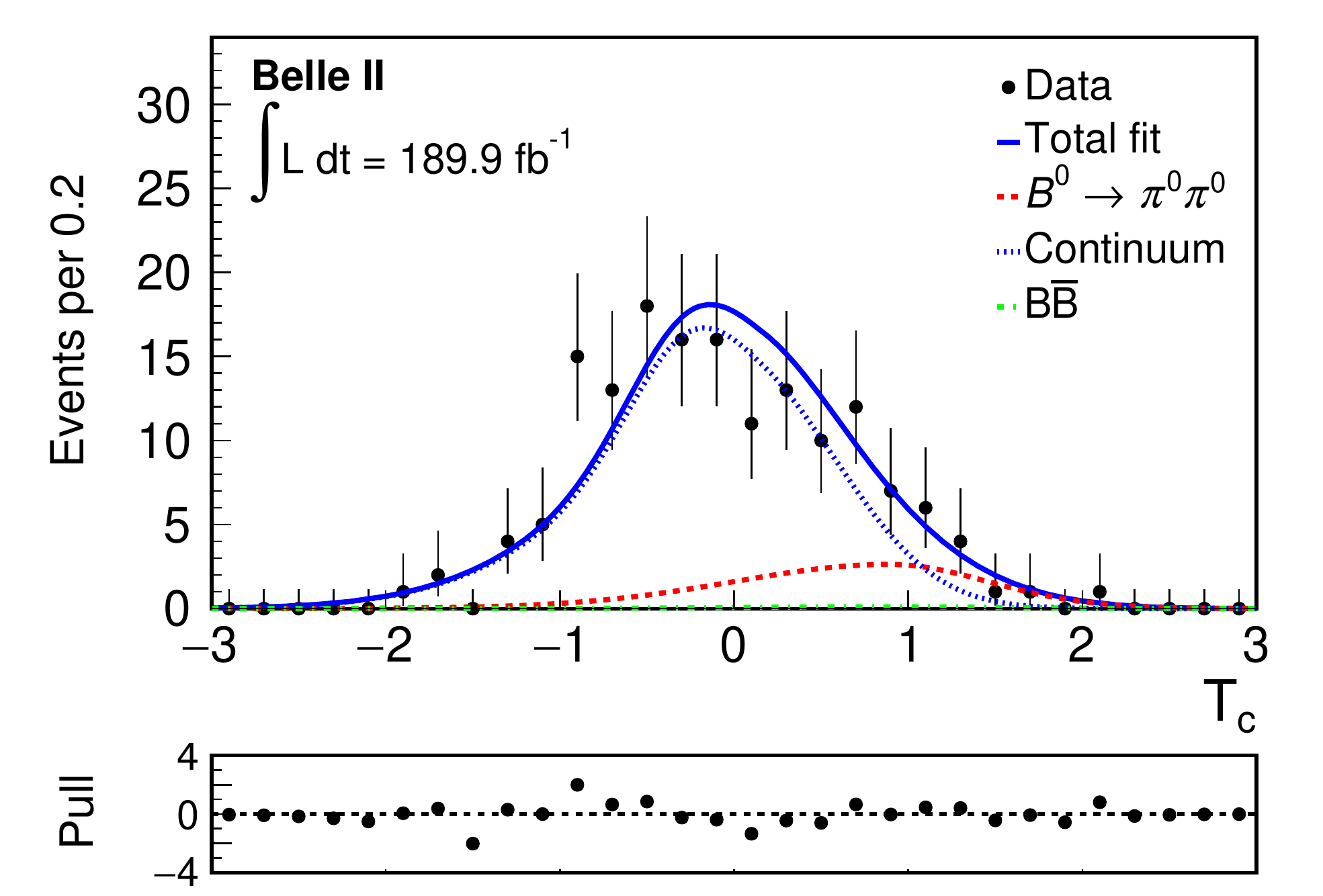}}
    \makebox[0.329\textwidth]{\includegraphics[clip, trim=0.75cm 0cm 0.2cm 0cm, width=0.333\textwidth]{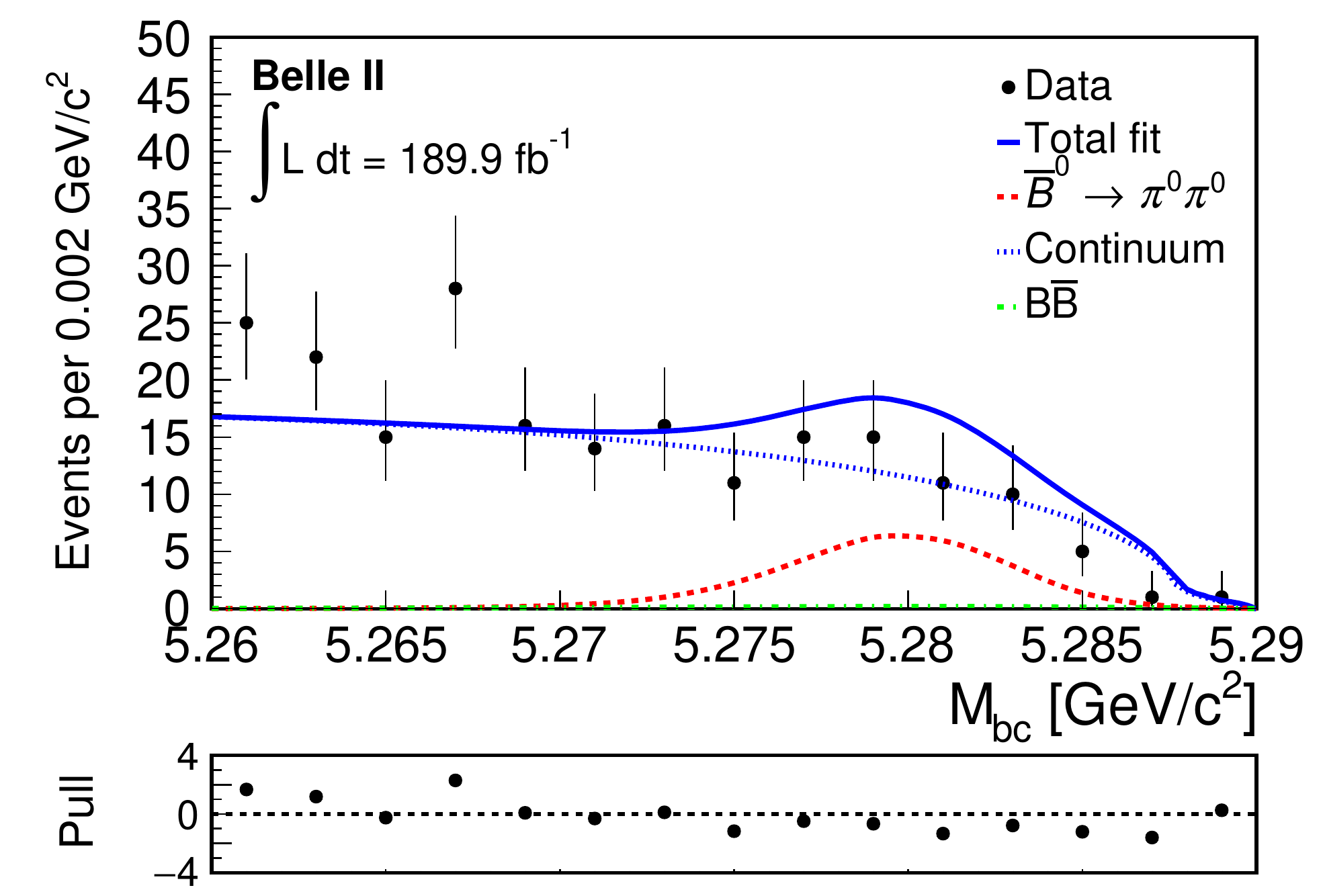}}
    \makebox[0.329\textwidth]{\includegraphics[clip, trim=0.75cm 0cm 0.2cm 0cm,width=0.333\textwidth]{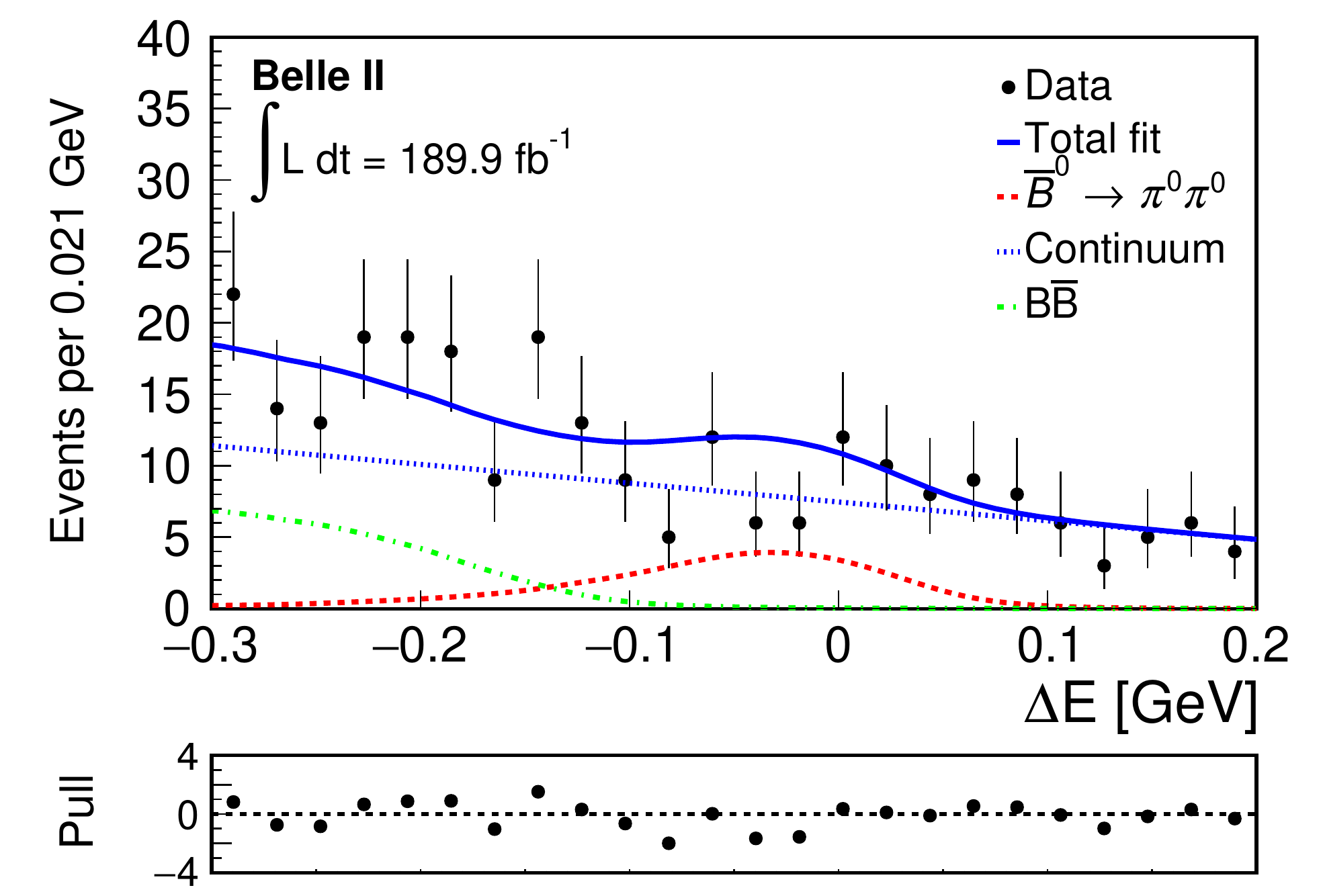}}    
    \makebox[0.329\textwidth]{\includegraphics[clip, trim=0.75cm 0cm 0.2cm 0cm,width=0.333\textwidth]{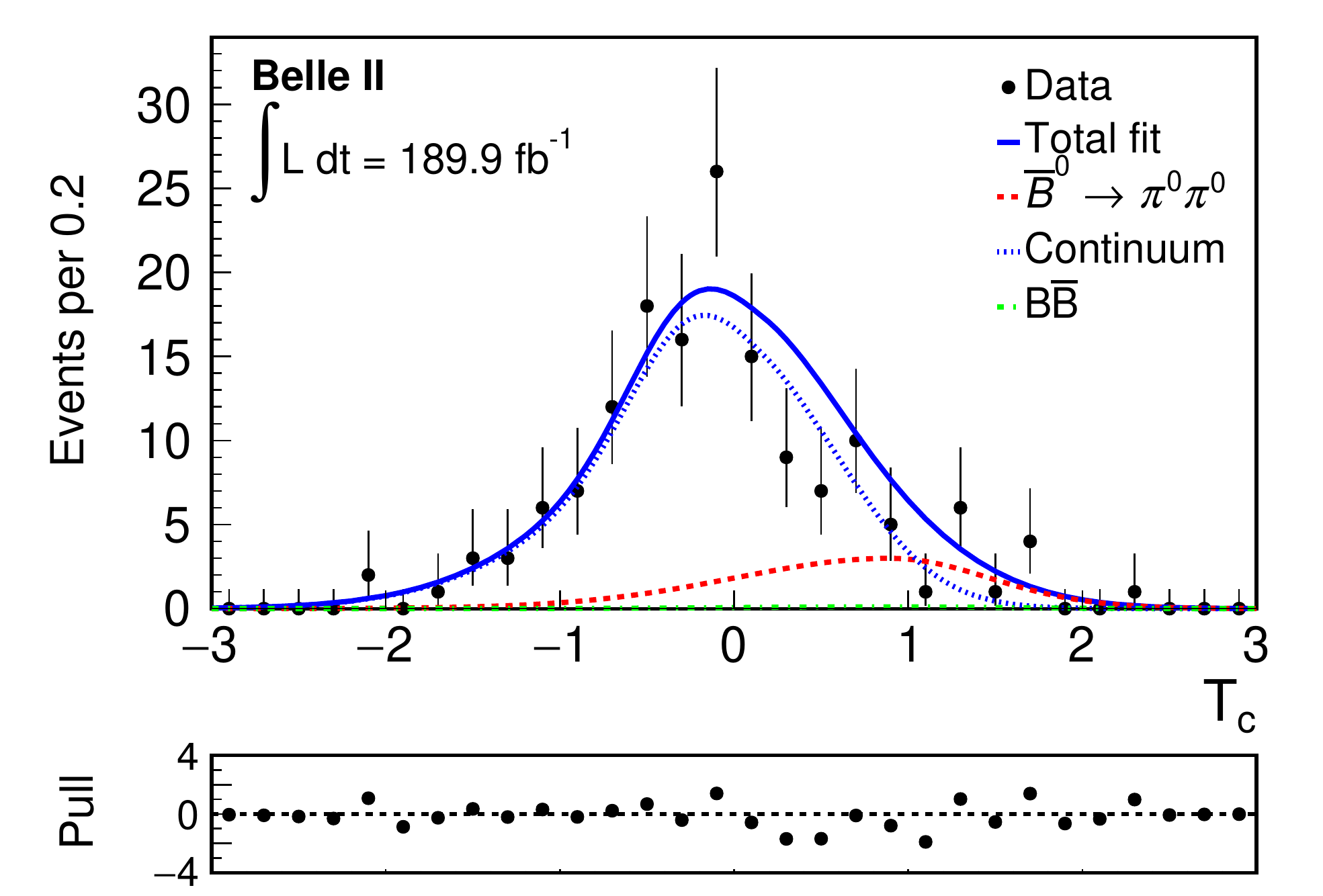}}
    \caption{Distributions of $\mbc$ (left), $\dele$ (middle), and $T_{c}$ (right) for the $\Bpipi$ candidates, for all seven $r$ bins combined, with positive (top) and negative (bottom) $q$ tags. The result of the fit to the data is shown as a solid blue curve. The fit components are shown as a red dashed curve (signal), blue dotted curve (continuum background), and green dash-dotted curve ($B\overline{B}$ background). The plots are signal-enhanced, which correspond to candidates with $5.275 < \mbc <5.285 \gevcc$, $-0.10 < \dele < 0.05 \gev$, and $0 < T_{c} < 3$. When the respective variable is displayed, the selections on that variable are not applied. The difference between observed and fit value divided by the uncertainty from the fit (pulls) are shown below each distribution.}
    \label{fig:ICHEP_fit_split}
\end{figure*}

We apply the fit described above to the 3177 selected \Bpipi candidate events. The signal yield, $\Acp$, and continuum yield are free to vary, while the $B\overline{B}$ yield is fixed to the expectation from simulations. We obtain a signal yield of $93 \pm 18$ events. Figure~\ref{fig:ICHEP_fit_split} shows the signal-enhanced projections of the fits to data, separately for positive and negative $q$ tags. The signal-enhanced region for the $\Bpipi$ signal decay is the same as that for the $B^0 \to \overline{D}{}^0(\to K^+ \pi^- \pi^0) \pi^0$ control mode and rejects approximately 96\% of the continuum background. To determine the signal significance, we convolve the statistical and additive systematic uncertainties and calculate the test statistic $2(\log{\mathcal{L}_m} - \log{\mathcal{L}_0}) = 32.0$ with two degrees of freedom, where $\log{\mathcal{L}_m}$ is the log-likelihood of the measured signal yield and $\log{\mathcal{L}_0}$ is determined by fixing the signal yield to zero. The second degree of freedom is lost due to $\Acp = 0$ when there is no signal. A total significance of 5.2 standard deviations is obtained.

The branching fraction is calculated using
\begin{equation}
    \mathcal{B}(\Bpipi) = \frac{N_{s}(1+f^{+-}/f^{00})}{2 \ \varepsilon \ N_{B\overline{B}} \ \mathcal{B}(\pi^0 \to \gamma \gamma)^2},
\end{equation}
where $N_{s}$ is the signal yield, $\varepsilon$ is the signal reconstruction and selection efficiency, $N_{B\overline{B}}$ is the number of $B\overline{B}$ pairs produced, $\mathcal{B}(\pi^0 \to \gamma \gamma)$~\cite{ParticleDataGroup:2022pth} is the $\piz \to \gamma \gamma$ branching fraction, and $f^{+-}/f^{00}$ is the ratio of the branching fractions for the decay of $\Y4S$ to $\BpBm$ and $\Bz \overline{B}{}^0$. The ratio $f^{+-}/f^{00}$ is determined to be $1.065 \pm 0.012 \pm 0.019 \pm 0.047$~\cite{Belle:2022hka}, where the first and second uncertainties are statistical and systematic, respectively, and the third uncertainty is due to the assumption of isospin symmetry in $B \to \jpsi(\to \ell\ell) K$, where $\ell = e$ or $\mu$. Inserting the values $N_{s} = 93 \pm 18$, $\varepsilon = (35.5 \pm 4.7)\%$, $N_{B\overline{B}} = (198.0 \pm 3.0) \times 10^{6}$, and $\mathcal{B}(\pi^0 \to \gamma \gamma) = (98.823 \pm 0.034) \%$~\cite{ParticleDataGroup:2022pth}, we obtain
\begin{equation}
    \mathcal{B}(\Bpipi) = (1.38 \pm 0.27 \pm 0.22) \times 10^{-6},
\end{equation}
where the first and second uncertainties are statistical and systematic (discussed below), respectively. The uncertainty in $\varepsilon$ is due to the systematic uncertainty associated with $\pi^0$ reconstruction and continuum classifier efficiency.

\begin{table}[!h]
\begin{tabular}{lccl}
\cline{1-3}
\noalign{\vskip\doublerulesep
         \vskip-\arrayrulewidth} \cline{1-3}
\textbf{Source}                & $\textbf{\BR (\%)}$ & $\textbf{\Acp}$ & \\ \cline{1-3}
$\pi^0$ reconstruction efficiency             & 11.6   &   n/a                             &  \\ 
Continuum parametrization                  & 7.4    & 0.02                              &  \\ 
Continuum classifier efficiency                 & 6.5         &  n/a                        &  \\ 
$1 + f^{+-}/f^{00}$        &     2.5         &          n/a                &  \\ 
Fixed $B\overline{B}$ background yield          & 2.3  & 0.01                              &  \\ 
Fixed signal $r$ bin fractions                  & 2.2    &  0.01 & \\   
Knowledge of the photon-energy scale                   & 2.0    & n/a                                &  \\ 
Assumption of independence of \dele from $r$ & 1.8 &      $< 0.01$                           &  \\ 
Number of $B\overline{B}$ meson pairs & 1.5  &         $< 0.01$                        &  \\ 
Choice of $(\mbc, \dele)$ signal model                  & 1.3  & 0.02                                &  \\ 
Fixed continuum $r$ bin fraction                 & 1.1    &  $< 0.01$                              &  \\ 

Branching fraction fit bias                & 1.0    &  n/a                              &  \\ 
Best candidate selection       & 0.2    &        $< 0.01$                       &  \\ 
Mistagging parameters               &  n/a   & 0.05                              &  \\ 
Potential non-zero $B\overline{B}$ background \Acp &   n/a   & 0.03                              &  \\
\Acp fit bias                    &  n/a     & 0.02                              &  \\ 
Continuum \qr asymmetry &  n/a    & 0.01                              &  \\

\cline{1-3}
Total                          &    16.2 &  0.07                           &  \\ \cline{1-3}
\noalign{\vskip\doublerulesep
         \vskip-\arrayrulewidth} \cline{1-3}
\end{tabular}
\caption{Summary of systematic uncertainties. The total is calculated by adding all systematic uncertainties in quadrature.}
\label{table:uncertainty}
\end{table}

The main sources of systematic uncertainties are listed in Table~\ref{table:uncertainty} and are evaluated as follows. A 3.4\% systematic uncertainty associated with the $\piz$ reconstruction efficiency is determined from data using the decays $D^{*-} \to \overline{D}{}^0(\to K^+ \pi^- \pi^0) \pi^-$ and $D^{*-} \to \overline{D}{}^0(\to K^+ \pi^-) \pi^-$, where the $\piz$ selection is identical to that of the signal. The $\piz$ reconstruction efficiency as a function of momentum is also measured using $\tau^-\to3\pi\piz\nu$ and $\tau^-\to3\pi\nu$ decays. A difference of 4.7\% in efficiency is observed between the measurement based on $D$ decays and the measurement based on $\tau$ leptons. This difference increases the systematic uncertainty for a total of 5.8\% per pion. The total systematic uncertainty associated with the $\pi^0$ reconstruction efficiency is then 11.6\%, as there are two pions and their errors are fully correlated.

The systematic uncertainty associated with the continuum parametrization accounts for the uncertainty in each of the data-driven continuum PDF parameters. The contribution of each parameter is determined by refitting on simulated data with the parameter used in the continuum PDF fluctuated by its one-standard-deviation uncertainties. All other continuum PDF parameters are correspondingly shifted according to their correlation with the fluctuated parameter. The systematic uncertainty is the sum in quadrature of the change in signal yield for each parameter. 

The systematic uncertainty associated with the efficiency of the continuum classifier is determined using $B^0 \to \overline{D}{}^0(\to K^+ \pi^- \pi^0) \pi^0$ decays. The efficiencies of the classifier selection in data and simulation are consistent within the statistical uncertainties. The overall statistical uncertainty is assigned as systematic uncertainty. The $f^{+-}/f^{00}$ systematic uncertainty combines the original systematic uncertainty and the uncertainty due to the assumption of isospin symmetry. The systematic uncertainty associated with the fixed $B\overline{B}$ background yield is determined by refitting on simulated data with the generated $B\overline{B}$ yield fluctuated by its one-standard-deviation uncertainties. The systematic uncertainty associated with the fixed signal fractions for $r$ bins is determined by refitting simulated data with the signal fractions fluctuated by their one-standard-deviation uncertainties. The systematic uncertainty is the sum in quadrature of the change in signal yield for each bin. A similar procedure to determine the systematic uncertainty associated with fixing the continuum fractions in the $r$ bins is also performed. The systematic uncertainty associated with the photon-energy corrections is determined by refitting on data with the values of the corrections fluctuated by their uncertainties. The largest change in yield is taken as the systematic uncertainty. The systematic uncertainty associated with the assumption of independence of \dele from $r$ is determined by refitting on simulated data with the $\dele$ slope for each $r$ bin separately estimated using large simulated samples. The procedure for estimating the uncertainty in the number of $B\overline{B}$ meson pairs is described in Ref.~\cite{Belle-II:2019usr}. The systematic uncertainty associated with the choice of $(\mbc, \dele)$ signal models is determined by refitting on simulated data with two uncorrelated Crystal Ball functions~\cite{Skwarnicki:1986xj}. A small bias in the calculated branching fraction due to the limitations of the PDFs used to model the data is included as a systematic uncertainty. The systematic uncertainty associated with a possible bias due to best candidate selection is determined by refitting on data with the best candidate randomly selected. The total systematic uncertainty is taken to be the sum in quadrature of the individual contributions (16.2\%).

The $\it CP$ asymmetry of $B^0 \to \pi^0 \pi^0$ decays is measured to be
\begin{equation}
    \Acp(\Bpipi) = 0.14 \pm 0.46 \pm 0.07.
\end{equation}
The main sources of systematic uncertainties are listed in Table~\ref{table:uncertainty} and are evaluated as follows. The systematic uncertainties for the continumm parameterization, fixed $B\overline{B}$ background yield, fixed signal $r$ bin fraction, choice of $(\mbc, \dele)$ signal model, and \Acp fit bias are evaluated as previously described. The systematic uncertainty due to the fixed values of $w_k$, $\Delta w_k$, and $\Delta \epsilon_k$ are determined by refitting simulated data with each parameter individually fluctuated by its one-standard-deviation uncertainties. The systematic uncertainty is the sum in quadrature of the change in \Acp for each parameter. The systematic uncertainty associated with bias due to potential non-zero $\Acp$ for the $B\overline{B}$ background is determined by refitting on simulated data with the generated \Acp for the two dominant $B\overline{B}$ backgrounds fluctuated by one standard deviation from their known values. The systematic uncertainty associated with the \qr asymmetry of the continuum is determined by refitting the data with the \qr asymmetry term of the continuum PDF fluctuated by its one-standard-deviation uncertainty. 

We average our results with previous measurements of $\mathcal{B}$ and \Acp for $B^0 \to \pi^0 \pi^0$ and use the isospin analysis in Ref.~\cite{Gronau1990}, along with previous measurements of the branching fractions and $\it CP$-asymmetry parameters for $B^0 \to \pi^+ \pi^-$ and $B^+ \to \pi^+ \pi^0$~\cite{ParticleDataGroup:2022pth}, and find that the $\phi_2$ exclusion interval at the 68\% confidence level increases by $1.0^{\circ}$, corresponding to a relative increase in precision of 1.4\%. Similarly, at the $95\%$ confidence level the exclusion interval increases by $1.3^{\circ}$, corresponding to a relative increase in precision of 2.0\%.

In conclusion, we measure the branching fraction and direct $\it CP$ asymmetry to be $\mathcal{B}(\Bpipi) = (1.38 \pm 0.27 \pm 0.22) \times 10^{-6}$ and $\Acp = 0.14 \pm 0.46 \pm 0.07$, respectively. These measurements agree with previous measurements~\cite{ParticleDataGroup:2022pth}. The branching fraction uncertainty is similar in size to those reported by the Babar and Belle collaboration, despite using a sample 2.4 and 4.0 times smaller, respectively. These improvements are due to a 60\% higher signal efficiency with approximately 40\% less background~\cite{Belle:2017lyb}. The higher efficiency and lower background result from improved photon timing, BDT-based photon selection, and data-driven continuum suppression. 

\input{acknowledgements}

\bibliographystyle{apsrev4-1}
\bibliography{belle2}

\end{document}

%% file: authors-orcid.tex
  \author{F.~Abudin{\'e}n\,\orcidlink{0000-0002-6737-3528}} 
  \author{I.~Adachi\,\orcidlink{0000-0003-2287-0173}} 
  \author{K.~Adamczyk\,\orcidlink{0000-0001-6208-0876}} 
  \author{L.~Aggarwal\,\orcidlink{0000-0002-0909-7537}} 
  \author{H.~Ahmed\,\orcidlink{0000-0003-3976-7498}} 
  \author{H.~Aihara\,\orcidlink{0000-0002-1907-5964}} 
  \author{N.~Akopov\,\orcidlink{0000-0002-4425-2096}} 
  \author{A.~Aloisio\,\orcidlink{0000-0002-3883-6693}} 
  \author{N.~Anh~Ky\,\orcidlink{0000-0003-0471-197X}} 
  \author{D.~M.~Asner\,\orcidlink{0000-0002-1586-5790}} 
% ...
\collaboration{The Belle II Collaboration}

%% file: acknowledgements.tex
This work, based on data collected using the Belle II detector, which was built and commissioned prior to March 2019, was supported by
Science Committee of the Republic of Armenia Grant No.~20TTCG-1C010;
Australian Research Council and research Grants
No.~DE220100462,
No.~DP180102629,
No.~DP170102389,
No.~DP170102204,
No.~DP150103061,
No.~FT130100303,
No.~FT130100018,
and
No.~FT120100745;
Austrian Federal Ministry of Education, Science and Research,
Austrian Science Fund
No.~P~31361-N36
and
No.~J4625-N,
and
Horizon 2020 ERC Starting Grant No.~947006 ``InterLeptons'';
Natural Sciences and Engineering Research Council of Canada, Compute Canada and CANARIE;
Chinese Academy of Sciences and research Grant No.~QYZDJ-SSW-SLH011,
National Natural Science Foundation of China and research Grants
No.~11521505,
No.~11575017,
No.~11675166,
No.~11761141009,
No.~11705209,
and
No.~11975076,
LiaoNing Revitalization Talents Program under Contract No.~XLYC1807135,
Shanghai Pujiang Program under Grant No.~18PJ1401000,
Shandong Provincial Natural Science Foundation Project~ZR2022JQ02,
and the CAS Center for Excellence in Particle Physics (CCEPP);
the Ministry of Education, Youth, and Sports of the Czech Republic under Contract No.~LTT17020 and
Charles University Grant No.~SVV 260448 and
the Czech Science Foundation Grant No.~22-18469S;
European Research Council, Seventh Framework PIEF-GA-2013-622527,
Horizon 2020 ERC-Advanced Grants No.~267104 and No.~884719,
Horizon 2020 ERC-Consolidator Grant No.~819127,
Horizon 2020 Marie Sklodowska-Curie Grant Agreement No.~700525 "NIOBE"
and
No.~101026516,
and
Horizon 2020 Marie Sklodowska-Curie RISE project JENNIFER2 Grant Agreement No.~822070 (European grants);
L'Institut National de Physique Nucl\'{e}aire et de Physique des Particules (IN2P3) du CNRS (France);
BMBF, DFG, HGF, MPG, and AvH Foundation (Germany);
Department of Atomic Energy under Project Identification No.~RTI 4002 and Department of Science and Technology (India);
Israel Science Foundation Grant No.~2476/17,
U.S.-Israel Binational Science Foundation Grant No.~2016113, and
Israel Ministry of Science Grant No.~3-16543;
Istituto Nazionale di Fisica Nucleare and the research grants BELLE2;
Japan Society for the Promotion of Science, Grant-in-Aid for Scientific Research Grants
No.~16H03968,
No.~16H03993,
No.~16H06492,
No.~16K05323,
No.~17H01133,
No.~17H05405,
No.~18K03621,
No.~18H03710,
No.~18H05226,
No.~19H00682, 
No.~22H00144,
No.~26220706,
and
No.~26400255,
the National Institute of Informatics, and Science Information NETwork 5 (SINET5), 
and
the Ministry of Education, Culture, Sports, Science, and Technology (MEXT) of Japan;  
National Research Foundation (NRF) of Korea Grants
No.~2016R1\-D1A1B\-02012900,
No.~2018R1\-A2B\-3003643,
No.~2018R1\-A6A1A\-06024970,
No.~2018R1\-D1A1B\-07047294,
No.~2019R1\-I1A3A\-01058933,
No.~2022R1\-A2C\-1003993,
and
No.~RS-2022-00197659,
Radiation Science Research Institute,
Foreign Large-size Research Facility Application Supporting project,
the Global Science Experimental Data Hub Center of the Korea Institute of Science and Technology Information
and
KREONET/GLORIAD;
Universiti Malaya RU grant, Akademi Sains Malaysia, and Ministry of Education Malaysia;
Frontiers of Science Program Contracts
No.~FOINS-296,
No.~CB-221329,
No.~CB-236394,
No.~CB-254409,
and
No.~CB-180023, and No.~SEP-CINVESTAV research Grant No.~237 (Mexico);
the Polish Ministry of Science and Higher Education and the National Science Center;
the Ministry of Science and Higher Education of the Russian Federation,
Agreement No.~14.W03.31.0026, and
the HSE University Basic Research Program, Moscow;
University of Tabuk research Grants
No.~S-0256-1438 and No.~S-0280-1439 (Saudi Arabia);
Slovenian Research Agency and research Grants
No.~J1-9124
and
No.~P1-0135;
Agencia Estatal de Investigacion, Spain
Grant No.~RYC2020-029875-I
and
Generalitat Valenciana, Spain
Grant No.~CIDEGENT/2018/020
Ministry of Science and Technology and research Grants
No.~MOST106-2112-M-002-005-MY3
and
No.~MOST107-2119-M-002-035-MY3,
and the Ministry of Education (Taiwan);
Thailand Center of Excellence in Physics;
TUBITAK ULAKBIM (Turkey);
National Research Foundation of Ukraine, project No.~2020.02/0257,
and
Ministry of Education and Science of Ukraine;
the U.S. National Science Foundation and research Grants
No.~PHY-1913789 
and
No.~PHY-2111604, 
and the U.S. Department of Energy and research Awards
No.~DE-AC06-76RLO1830, 
No.~DE-SC0007983, 
No.~DE-SC0009824, 
No.~DE-SC0009973, 
No.~DE-SC0010007, 
No.~DE-SC0010073, 
No.~DE-SC0010118, 
No.~DE-SC0010504, 
No.~DE-SC0011784, 
No.~DE-SC0012704, 
No.~DE-SC0019230, 
No.~DE-SC0021274, 
No.~DE-SC0022350; 
and
the Vietnam Academy of Science and Technology (VAST) under Grant No.~DL0000.05/21-23.

These acknowledgements are not to be interpreted as an endorsement of any statement made
by any of our institutes, funding agencies, governments, or their representatives.

We thank the SuperKEKB team for delivering high-luminosity collisions;
the KEK cryogenics group for the efficient operation of the detector solenoid magnet;
the KEK computer group and the NII for on-site computing support and SINET6 network support;
and the raw-data centers at BNL, DESY, GridKa, IN2P3, INFN, and the University of Victoria for offsite computing support.